\documentclass[aps,%
reprint,
prl,
groupedaddress]{revtex4-2}
\usepackage{color}
\usepackage{hyperref}
\usepackage{float}
\usepackage{amsmath,amsfonts,amssymb,epsfig,graphicx}
\usepackage{slashed} 

\newcommand{\MGMCatNLO}{MadGraph5\_aMC@NLO} 

\newcommand{\pt}{\ensuremath{p_\mathrm{T}}}

\newcommand{\PZ}{\ensuremath{\mathrm{Z}}}

\newcommand{\PW}{\ensuremath{\mathrm{W}}}

\newcommand{\mup}{\ensuremath{\mu^+}}
\newcommand{\mum}{\ensuremath{\mu^-}}
\newcommand{\elp}{\ensuremath{\mathrm{e}^+}}
\newcommand{\elm}{\ensuremath{\mathrm{e}^-}}
\newcommand{\nue}{\ensuremath{\nu_{\mathrm{e}}}}
\newcommand{\nuae}{\ensuremath{\tilde{\nu}_{\mathrm{e}}}}
\newcommand{\num}{\ensuremath{\nu_{\mu}}}
\newcommand{\nuam}{\ensuremath{\tilde{\nu}_{\mathrm{\mu}}}}

\newcommand{\fbinv}{\mbox{\ensuremath{~\mathrm{fb^{-1}}}}}
\newcommand{\al}{{\it et al. }}
\graphicspath{{./pic/}}
\begin{document}

\title{The physics case for a neutrino lepton collider in light of the CDF W mass measurement}


\author{Tianyi \surname{Yang}$^{1}$}

\author{Sitian \surname{Qian}$^{1}$}

\author{Sen \surname{Deng}$^{1}$}

\author{Jie \surname{Xiao}$^{1}$}

\author{Leyun \surname{Gao}$^{1}$}

\author{Andrew Michael \surname{Levin}$^{1}$}

\author{Qiang \surname{Li}$\ast$$^{1}$}

\author{Meng \surname{Lu}$^{2}$}

\author{Zhengyun \surname{You}$^{2}$}

\affiliation{1. State Key Laboratory of Nuclear Physics and Technology, School of Physics, Peking University, Beijing, 100871, China}
\affiliation{2. School of Physics, Sun Yat-Sen University, Guangzhou 510275, China}

\begin{abstract}
We propose a neutrino lepton collider where the neutrino beam is generated from TeV scale muon decays. Such a device would allow for a precise measurement of the W mass based on single W production $\nu\,l \rightarrow \PW^{(*)}$. Although it is challenging to achieve high instantaneous luminosity with such a collider, we find that a total luminosity of 0.1\fbinv can already yield competitive physics results. In addition to a W mass measurement, a rich variety of physics goals could be achieved with such a collider, including W boson precision measurements, heavy leptophilic gauge boson searches, and anomalous $\PZ\nu\nu$ coupling searches. A neutrino lepton collider is both a novel idea in itself, and may also be a useful intermediate step, with less muon cooling required, towards the muon-muon collider already being pursued by the energy frontier community. A neutrino neutrino or neutrino proton collider may also be interesting future options for the high energy frontier.
\end{abstract}

\maketitle

\section{Introduction.}
\label{introduction}

In recent years, we have witnessed several significant anomalies or hints of possible new physics beyond the Standard Model (SM).  First, the LHCb Collaboration, in a test of lepton flavour universality using $B^+ \rightarrow K^+ \ell^+ \ell^-$, reports a measurement that deviates by 3.1 standard deviations from the Standard Model (SM) prediction~\cite{Rk1}. Second, the latest result from the Muon g-2 Experiment at Fermilab has pushed the world average of the muon anomalous magnetic moment measurements to 4.2 standard deviations away from the SM prediction~\cite{gminus2}. Most recently, the CDF II collaboration~\cite{CDF:2022hxs} has reported a measurement of the $W$ gauge boson mass, $M_W^{\rm CDF} = 80.433 \pm 0.009 \,\, {\rm GeV}$, which is $7.2\sigma$ deviations away from the SM prediction of $M_W^{\rm SM} = 80.357 \pm 0.006 \,\, {\rm GeV}$ \cite{RPP}.  Numerous theoretical studies, e.g. Ref.~\cite{wmasspheno}, attempt to accommodate these anomalies, which may or may not require a modification of the SM.

These anomalies have also stimulated research and development for future experimental facilities. In the next two decades, the LHC and the High-Luminosity LHC (HL-LHC) will continue exploring the SM and searching for physics beyond that. Beyond the HL-LHC, there are quite a lot of proposals for the next generation collider for the purpose of Higgs boson related measurements, among which the lepton colliders are in the majority. The promising proposals include a linear or circular electron-positron collider~\cite{ILC, FCC, CEPC, CLIC} or a muon collider~\cite{Muc0,Muc1,Muc2,Muc3,Muc4}. Other options include an electron-muon collider~\cite{Lu:2020dkx},  a muon-proton collider~\cite{Cheung:2021iev}, or a muon-ion collider~\cite{Acosta:2021qpx}.

The difficulty of performing a $\PW$ mass measurement~\cite{CDF:2022hxs} lies in the fact that for $\PW\rightarrow l\nu$ ($l$=e or $\mu$), the neutrino escapes detectors, and thus one can reconstruct only the transverse mass instead of the invariant mass of the $\PW\rightarrow l\nu$ system (while $\PW\rightarrow qq$ is even more difficult because it must rely on hadron calorimeter). Both the transverse mass and lepton momentum are too difficult to model and calibrate well enough to achieve any vast improvement on the mass measurement using hadron colliders. However, if a collision beam of neutrinos could be created, one could then collect a clean sample of single W boson production ($\nu\,l \rightarrow \PW^{(*)}$), and possibly extract the W mass and width with limited integrated luminosity.

Several related neutrino scattering experiments have been proposed in the last few decades, including NuTeV~\cite{NuTeV:2001whx}, NuMAX~\cite{Delahaye:2018yfq},  NuSOnG~\cite{NuSOnG:2008weg}, and nuSTORM~\cite{nuSTORM:2012jbd}. Their motivations include, e.g., making precision neutrino interaction cross section measurements, or searching for neutrino related non-SM physics.  However, a head-on neutrino lepton collider at the 100 GeV scale is proposed in this letter for the first time, with rich physics potential discussed below.

\section{Experimental Setup}
\label{beam}

\begin{figure}
    \centering
    \includegraphics[width=.9\columnwidth]{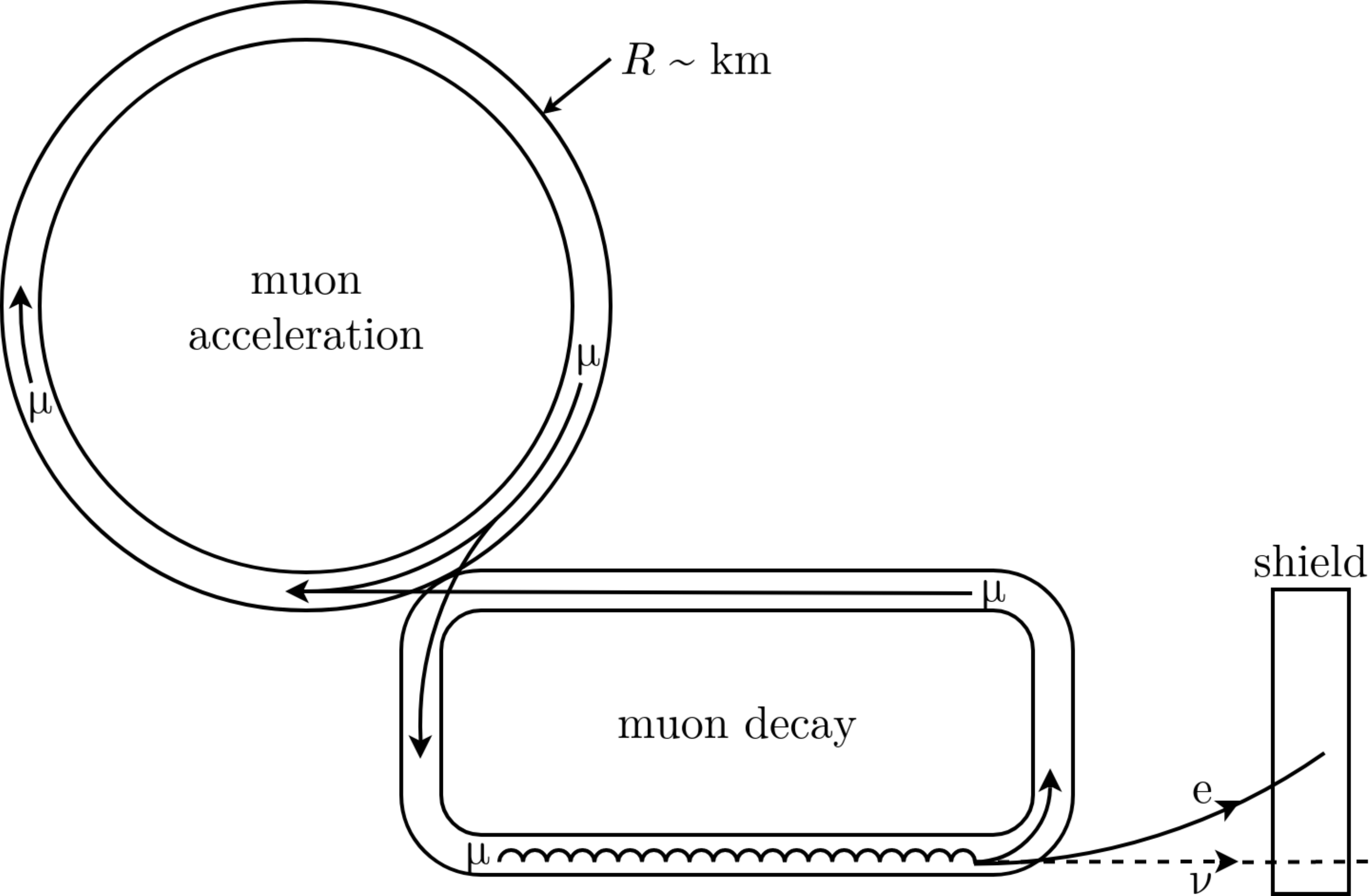}
    \caption{An illustration of the proposed neutrino beam and collider. Muons are accelerated in the circular section in the upper left, and then extracted into the rectangular section in the lower right. In one of the long edges of the rectangle, the neutrinos emitted from the muon decays are formed into a collimated beam. A small modulation of the muon decay angle through vertical bending, symbolized by the squiggly line, is used to focus the neutrino beam.}
    \label{fig:design}
\end{figure}

\begin{figure}
    \centering
    \includegraphics[width=0.4\textwidth]{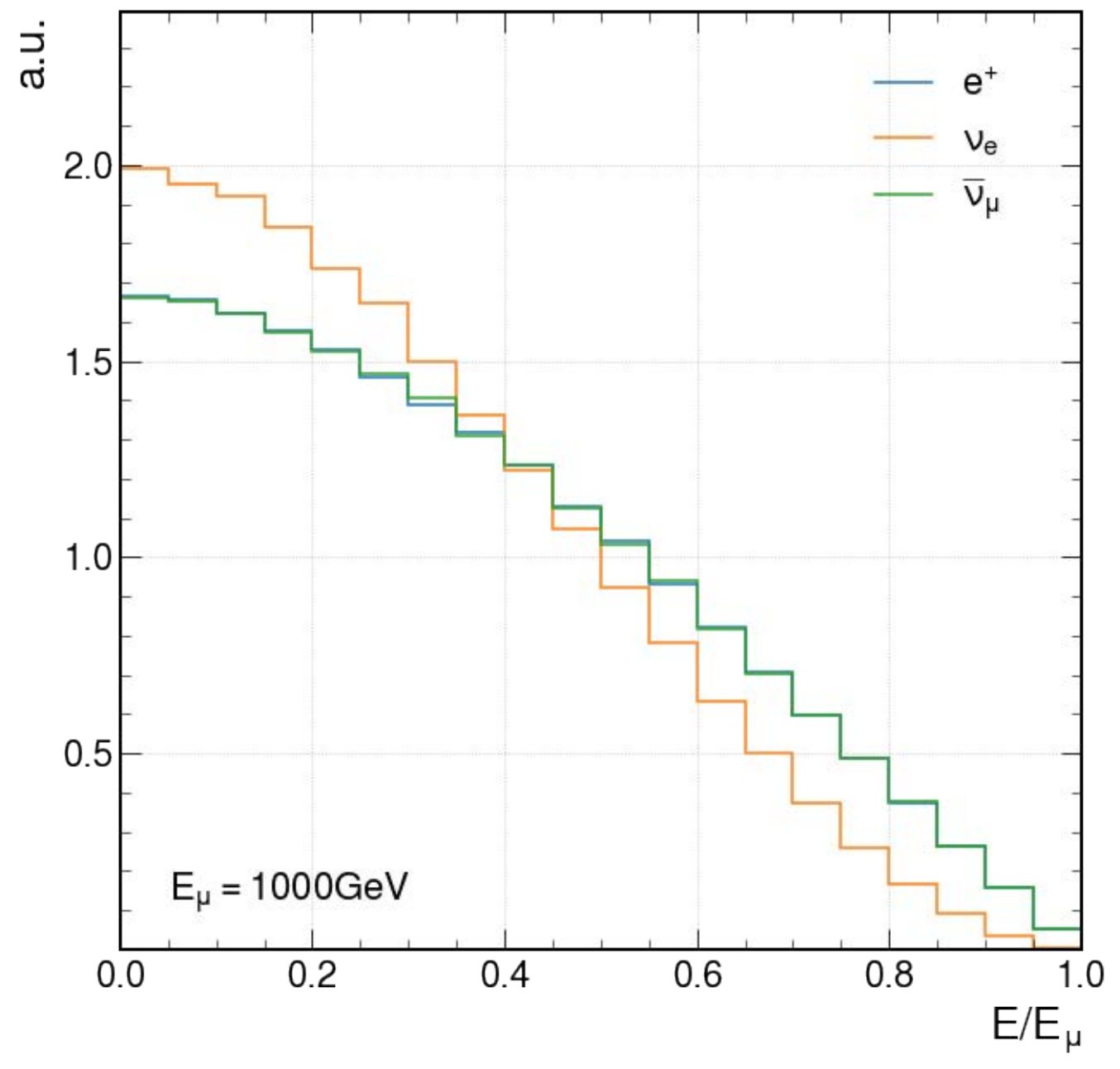}
    \includegraphics[width=0.4\textwidth]{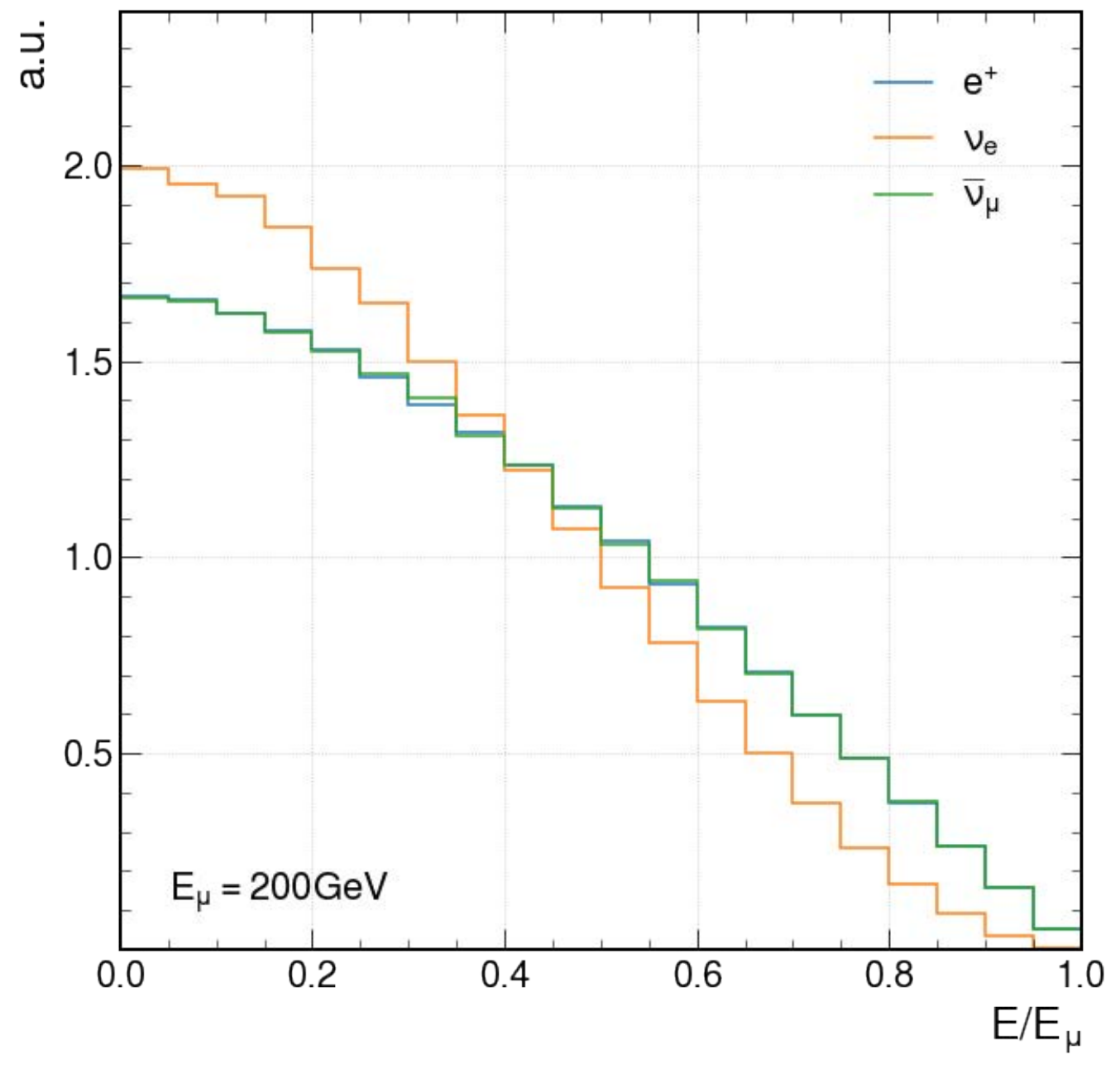}
    \caption{Energy fraction distributions of products emitted from 200 GeV and 1000 GeV muon beams.}
    \label{fig:Eneu}
\end{figure}

\begin{figure}
    \centering
    \includegraphics[width=0.4\textwidth]{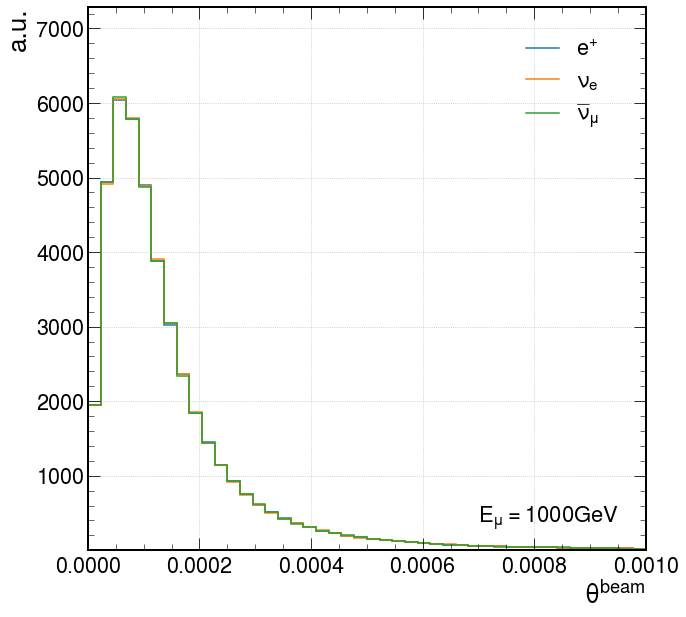}
    \includegraphics[width=0.4\textwidth]{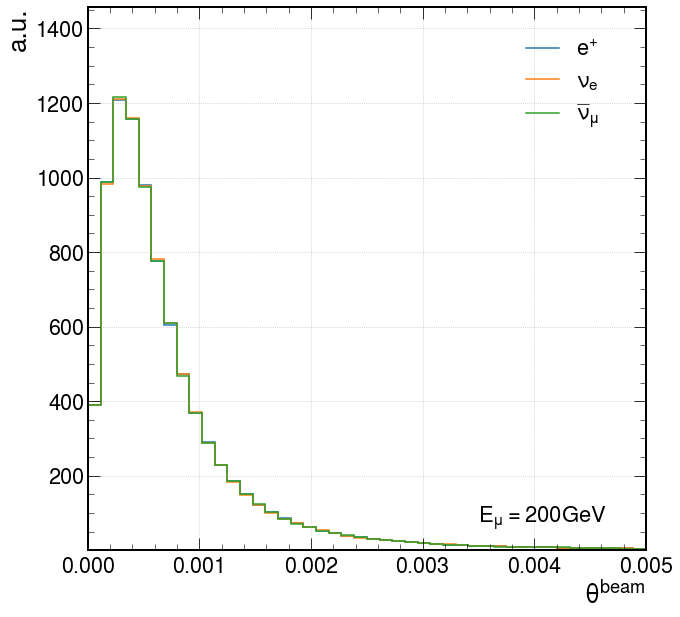}
    \caption{$\theta$ distributions of products emitted from 200 GeV and 1000 GeV muon beams.}
    \label{fig:Thetaneu}
\end{figure}

Taking a TeV scale $\mup\rightarrow  \elp\nue\nuam$ beam as an example. Fig.~\ref{fig:Eneu} shows the distributions of muon decay products' energy from a muon beam with energy at 200 GeV and 1 TeV. As the decay angle $\theta$ goes like $\theta\sim 10^{-4}/{\rm E(TeV)}$, the muon decay products will be more collimated with increasing beam energy~\cite{King:1999kx}. To prove this, Fig.~\ref{fig:Thetaneu} shows the distributions of theta of muon decay products. 

An illustration of the proposed neutrino beam and collider is shown in Fig.~\ref{fig:design}. The muon beam is accelerated in the circular section in the upper left and  then extracted into the rectangular section in the lower right. During each cycle, the beam will be squeezed due to Lorentz contraction and then pass through arc ($L_c$) and linear sections ($L_l$), emitting bunches of collimated neutrinos from the muon decays. The electrons from the muon decays can either be shielded or used for energy calibration through collision with positrons from the other side. Using the two rings instead of one ring here allows for more flexibility to accommodate crowded bunches with different time or space gaps.

The instantaneous luminosity of a neutrino lepton collider would be limited by two main factors: 1) the intensity of the neutrino beam compared with the incoming muon beam is suppressed by roughly $L_l/L_c\sim 0.1$, i.e., the fraction of the collider ring circumference occupied by the production straight section~\cite{King:1999kx}, 2) the neutrino beam spread, which may still be kept at 10 to 100 microns at the interaction point, by applying a small modulation on muon decay angle through vertical bending to achieve more focused neutrino beam~\cite{vbending}.

In more details, by using the formula for the instantaneous luminosity,
\begin{align}
  {\cal L} = {N_{\rm beam 1} N_{\rm beam 2} \over 4 \pi \sigma_x \sigma_y} f_{\rm rep},
\end{align}
where $f_{\rm rep}$ is the rate of collisions and is typically 100 kHz (40 MHz) for lepton colliders (hadron colliders), and $N_{\rm beam 1,2}$ are the number of particles in each bunch which can be taken as $\sim 10^{11}\text{--}10^{12}$~\cite{FCC:2018evy}, $\sigma_x$ and $\sigma_y$ are the beam sizes. Take the LHC as an example, with $f_{\rm rep}=40$\,MHz, $\sigma_{x,y}=16$ microns, and $N_{\rm beam 1,2}=10^{11}$, one can get $ {\cal L}=10^{34}$ cm$^{-2}$s$^{-1}$. As for TeV muon colliders~\cite{Bossi:2020yne,Delahaye:2019omf}, with $f_{\rm rep}=100$\,KHz, $\sigma_{x,y}\lesssim 10$ microns, and $N_{\rm beam 1,2}=10^{12}$, then $ {\cal L}=10^{33}\text{--}10^{34}$ cm$^{-2}$s$^{-1}$. As for the neutrino neutrino collisions discussed above, there are further suppression factors from linear over arc ratio ($L_l/L_c\sim 1/5$) with the exact value depending on the realistic design as shown in Fig.~\ref{fig:design}, and the neutrino beam spread which can be around 1000 microns for $L_l\sim$ 10 to 100 meters. Taking all these into account, a realistic instantaneous luminosity for neutrino neutrino collisions can reach around $ {\cal L}=10^{29-30}$ cm$^{-2}$s$^{-1}$ level.

On the other hand, the lepton beam from the other collision side is of lower energy with a few GeV, and the quality can be improved by many high-current high-frequency techniques. We assume here that for a neutrino electron collider with the neutrinos emitted by TeV scale muon beams, and electron energies around 5 GeV, the instantaneous luminosity can be increased furthermore, and in the following study, we assume the integrated luminosity to be around 1-10\fbinv in 10 years.

\section{Physics Potential}
\label{amm}

With a TeV scale $\mup\rightarrow  \elp\nue\nuam$ beam, if the collision beams from the other side are of $\elm$, $\elp$ and $\mum$, respectively, some of the main physics processes can be shown as below:
\begin{align}
&\elp \elm \rightarrow \PZ^{0(*)},\,\,\, \nue \elm \rightarrow \nue \elm,\,\,\, \nuam \elm \rightarrow \nuam \elm, \\
&\nue \elp \rightarrow \PW^{+(*)},\,\,\, \nuam \elp \rightarrow \nuam \elp,\,\,\, \nuam \elp \rightarrow \nuae \mup, \\
&\nuam \mum \rightarrow \PW^{-(*)},\,\,\, \nue \mum \rightarrow \nue \mum,\,\,\, \nue \mum \rightarrow \elm \num .
    \label{eq:mupdecay}
\end{align}

\begin{figure}
    \centering
    \includegraphics[width=0.42\textwidth]{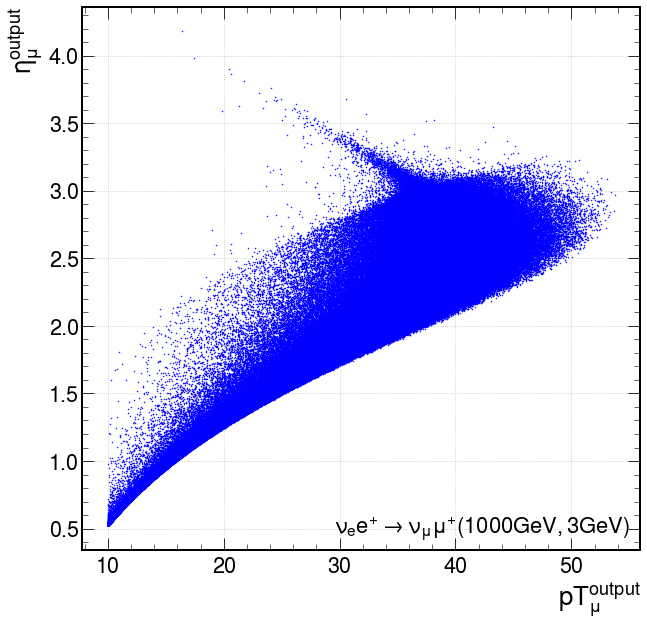}
    \includegraphics[width=0.42\textwidth]{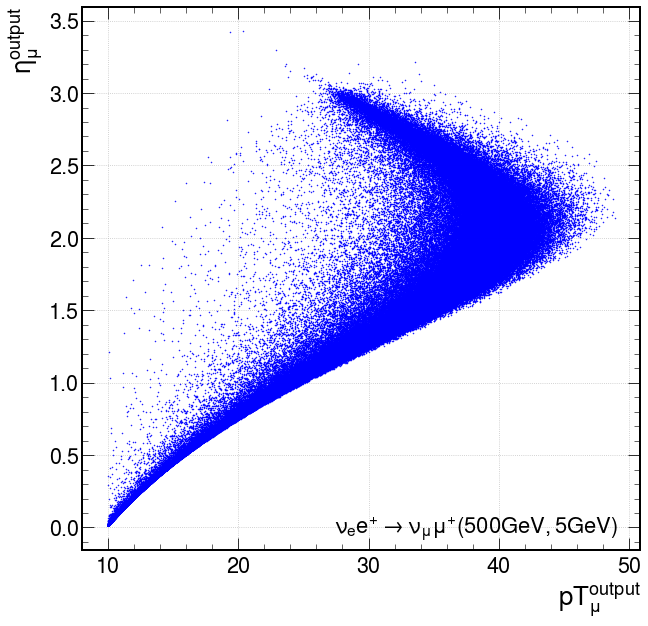}
    \caption{Scatter plot between the outgoing muon $\eta$ and $\pt$ from $\nue \elp \rightarrow \PW^{+(*)} \rightarrow \num\mup$ at two collision scenarios: a neutrino beam arising from a 1000 (500) GeV muon beam, and a 3 (5) GeV positron beam.}
    \label{fig:eta_pt_emu}
\end{figure}

We are especially interested in $\nue \elp \rightarrow \PW^{+(*)}$, which has a cross section that depends on $M_W$. To simulate this process, we implement the neutrino energy fraction function shown above (Fig.~\ref{fig:Eneu}) in \MGMCatNLO~\cite{Alwall:2014hca}. We simulated $\nue \elp \rightarrow \PW^{+(*)} \rightarrow \num\mup$ for two beam energy scenarios: a neutrino beam arising from a 1000 (500) GeV muon beam, and a 3 (5) GeV positron beam. Fig.~\ref{fig:eta_pt_emu} shows the $\pt$ and $\eta$ scatter plots of the outgoing muon. Accordingly, we require the final state muon to satisfy $\pt>10$\,GeV and $0<\eta<3.0$. 

\begin{figure}
    \centering
    \includegraphics[width=0.42\textwidth]{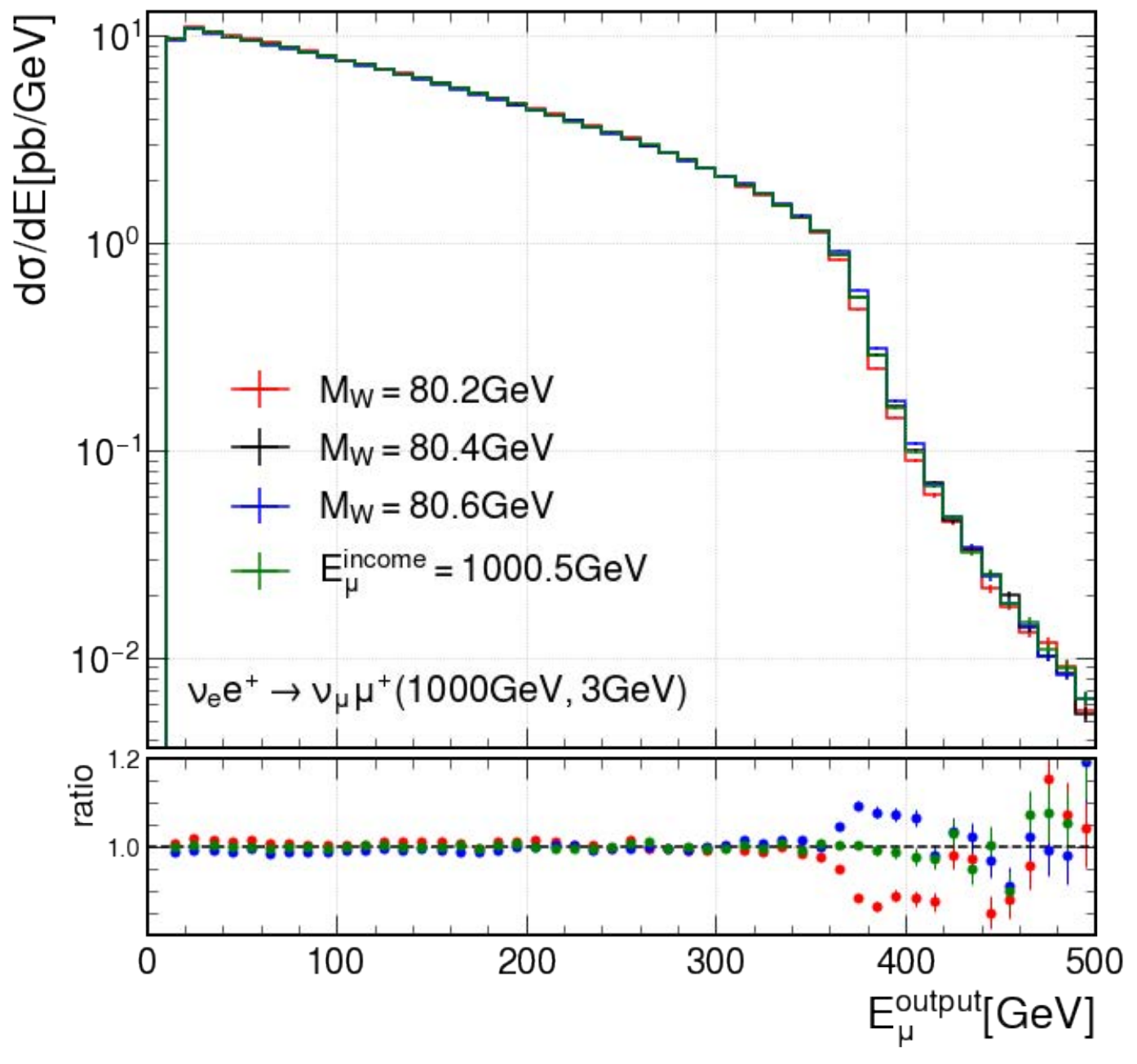}
    \includegraphics[width=0.42\textwidth]{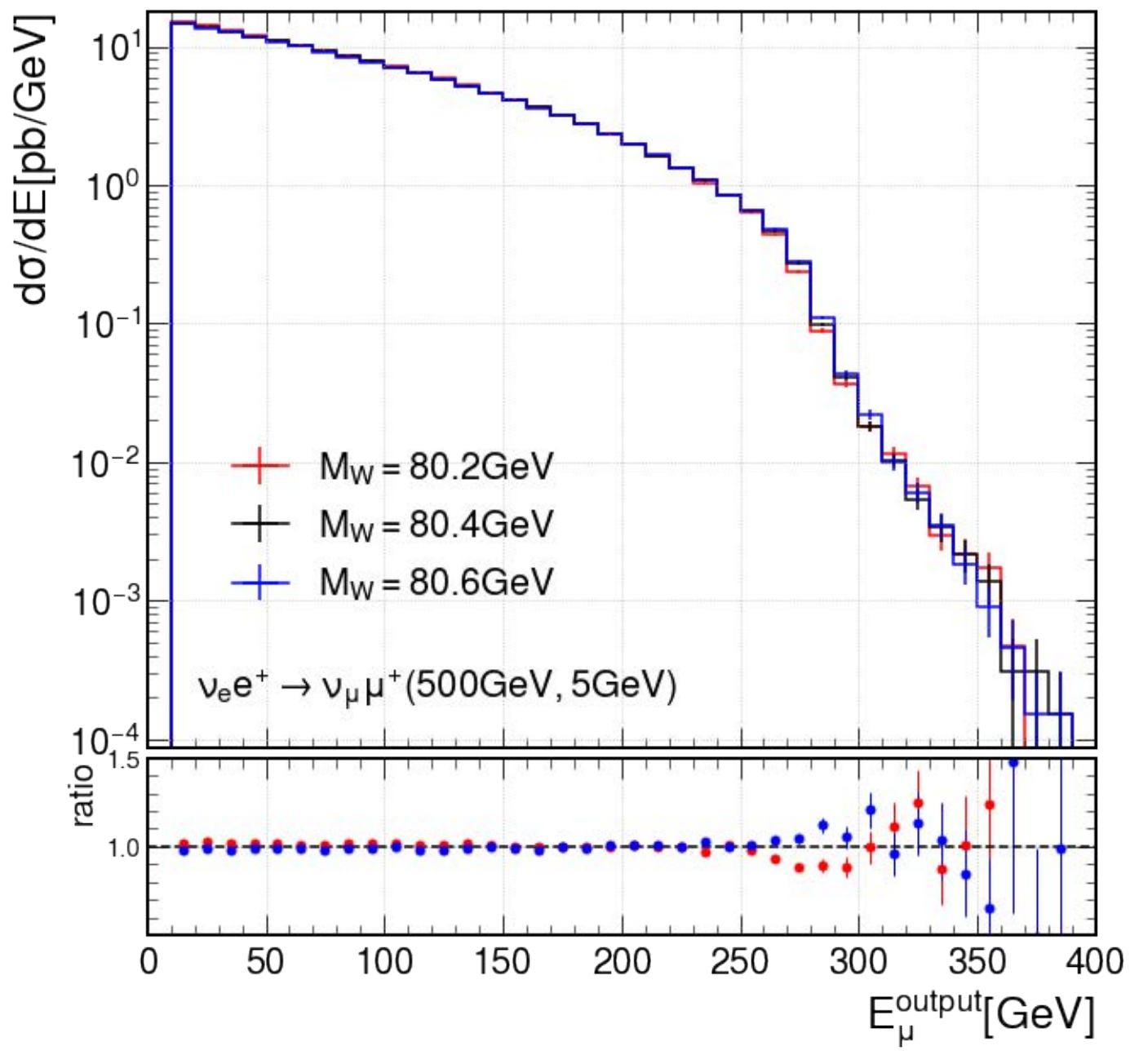}
    \caption{Distributions of the outgoing muon energy from $\nue \elp \rightarrow \PW^{+(*)} \rightarrow \num\mup$ at two collision scenarios: a neutrino beam arising from a 1000 (500) GeV muon beam, and a 3 (5) GeV positron beam. Clearly visible differences are seen between the $M_W=80.2$, $80.4$ and $80.6$\,GeV cases. The figure above also shows the energy comparison of output muon when the income muon energy of 1000 GeV varies by 0.5 GeV at $M_W=80.4\mathrm{GeV}$. Ratios are defined as distributions in other colors divided by the distribution in black ($E_{\mu}^{\mathrm{income}}=$1000 (500) GeV and $M_W=80.4\mathrm{GeV}$). Error bars include only the statistical errors.}
    \label{fig:emu}
\end{figure}

In Fig.~\ref{fig:emu}, we show the distributions of outgoing muon energy corresponding to the three mass hypotheses $M_W=80.2$, $80.4$, and $80.6$\,GeV, respectively. One can see clear differences between the different cases in the high end of the spectrum, with higher $\PW$ masses showing more high energy muons. Note that for positron beam with 3 (5) GeV, the energy of the incoming neutrino needed to reach the $M_W$ threshold is around 400-500 (200-300) GeV. This, together with the neutrino energy distribution function, explain the kinks in the figure. 

If the requirement $\pt>40$\,GeV is added, the cross sections at the [1000, 3] GeV neutrino electron collider with $M_W=80.4\, (80.41)$ are 166.2 (167.6) pb. Based on a simple counting experiment, a 10 MeV accuracy on $M_W$ can be achieved with an integrated luminosity of only 0.1\fbinv. In this calculation, we only consider W decays into muon. If we also include hadronic decay channels, and perform a more complicated shape analysis, the integrated luminosity needed to reach 10 MeV accuracy should be far less than 0.1\fbinv. A detailed systematic study is beyond the scope in this paper, however, we examined two possible sources. First, we found background contamination from $\nue \elp \rightarrow \nue \elp \PZ$ to be negligible. Second, we varied the incoming muon and electron beam energy by 0.5 GeV and 10 MeV, respectively, which are quite conservative following refs.~\cite{deBlas:2022aow} and ~\cite{Blondel:2019jmp}. We found that the cross sections changed by about 0.6 pb for both variations. This uncertainty could be mitigated by using the shape of the outgoing muon energy, by scanning different incoming beam energies, or by calibrating the incoming muon beam energy with the electron decay products.

Other potential physics results from such a neutrino lepton collider include a search for leptophilic gauge bosons~\cite{Buras:2021btx}, and studies of neutrino scattering processes, e.g., $\nue \elm \rightarrow \nue \elm$ and $\nue \mum \rightarrow \nue \mum$, which can be used to probe the $\PZ\nu\nu$ couplings ~\cite{NuTeV:2001whx,Davidson:2001ji}. For a neutrino lepton collider with neutrinos from a 1 TeV muon beam and a 5 (20) GeV electron beam, the scattering cross section is around 6 (20) pb. Thus one can expect to accumulate a million events similar to NuTeV~\cite{NuTeV:2001whx}, while the final states populated at higher energy with 10-100\fbinv of data already. Finally, this novel device may also shed light on the neutrino mixing matrix, the PMNS matrix, analogously to the B factories and CKM measurements. 

\section{Discussions}
\label{discussions}

In light of the recent W mass anomaly from the CDF collaboration, we propose a neutrino lepton collider that uses a highly collimated neutrino beam from TeV scale muon decays. Although it is quite challenging to achieve high instantaneous luminosity, due to limitations on the intensity and quality of a neutrino beam from muon decays, we find that a total luminosity of 0.1\fbinv is already sufficient to produce competitive physics results. We demonstrate that by performing a simple analysis on the $\nu\,l \rightarrow \PW^{(*)}$ process, a 10 MeV accuracy on $M_W$ can be achieved with an integrated luminosity of 0.5\fbinv. If we were to include hadronic decay channels and/or perform a shape analysis, the integrated luminosity needed to achieve 10 MeV accuracy would likely be less than 0.1\fbinv. Our proposed neutrino lepton collider would share some technological synergies with the muon collider being pursued by the energy frontier community, and so could be considered an intermediate step or a by-product of that effort. It may also require less cooling of the muon beams. Other neutrino collider concepts, such as a neutrino neutrino or a neutrino proton collider, may also be interesting future options for the high energy frontier.

\appendix

\begin{acknowledgments}
This work is supported in part by the National Natural Science Foundation of China under Grants No. 12150005, No. 12075004 and No. 12061141002, by MOST under grant No. 2018YFA0403900.

The datasets used and/or analysed during the current study available from the corresponding author on reasonable request.
\end{acknowledgments}


\bibliographystyle{ieeetr}
\bibliography{h}
\end{document}